\documentclass[aps,twocolumn,groupedaddress,floatfix,amsmath,amssymb,pra,longbibliography]{revtex4-1}
\usepackage{amsmath}
\usepackage{amssymb}
\usepackage{graphicx}
\usepackage{textcomp}
\usepackage{bm} 
\usepackage{braket} 
\usepackage{longtable}
\usepackage[usenames,dvipsnames]{color}

\begin{document}

%\linenumbers
\title{Slow-light enhanced frequency combs and dissipative Kerr solitons in silicon coupled-ring microresonators in the telecom band}

\author{L. Marti}
\affiliation{Institute of Theoretical Physics, \'Ecole Polytechnique F\'ed\'erale de Lausanne (EPFL), CH-1015 Lausanne, Switzerland}
\author{J. P. Vasco}
\affiliation{Institute of Theoretical Physics, \'Ecole Polytechnique F\'ed\'erale de Lausanne (EPFL), CH-1015 Lausanne, Switzerland}
\author{V. Savona}
\affiliation{Institute of Theoretical Physics, \'Ecole Polytechnique F\'ed\'erale de Lausanne (EPFL), CH-1015 Lausanne, Switzerland}

%\pacs{xxxx, xxxx, xxxx}                                         
    %\date{\today}
    
\begin{abstract} 
We propose a system of coupled microring resonators for the generation frequency combs and dissipative Kerr solitons in silicon at telecommunication frequencies. By taking advantage of structural slow-light, the effective non-linearity of the material is enhanced, thus relaxing the requirement of ultra-high quality factors that currently poses a major obstacle to the realization of silicon comb devices. We demonstrate a variety of frequency comb solutions characterized by threshold power in the 10-milliwatt range and a small footprint of $0.1$~mm$^2$, and study their robustness to structural disorder. The results open the way to the realization of low-power compact comb devices in silicon at the telecom band.  
\end{abstract} 

\maketitle

\section{Introduction}

Kerr frequency combs in microresonators have been the object of intense research during the last decade, due to their wide range of applications in science and engineering \cite{chembo7,baumann,gaeta5,vahala2}. They have brought about significant advances in several areas of sensing and communications, such as light detection and ranging (LIDAR) \cite{koos2}, optical atomic clocks \cite{vahala3}, exoplanet exploration \cite{herr}, optical frequency synthesis \cite{scott} and high-resolution spectroscopy \cite{swann,hansch}. Kerr frequency combs originate from an interplay between the Kerr non-linearity and the frequency dispersion. More precisely, a comb is generated when the free spectral range (FSR) of the resonator increases with frequency (anomalous dispersion), thereby causing multiple parametric resonant four-wave mixing (FWM) processes, leading to a frequency comb of evenly spaced emission lines \cite{kippenberg5,kippenberg3}. When all the targeted resonant frequencies of the resonator participate in the parametric process, the non-linear dynamics may give rise to a comb with a single FSR frequency spacing, known as dissipative Kerr solitons (DKS). A DKS is the result of two balances: the one between Kerr non-linearity and dispersion, which stabilizes their spectral shape, and the one between linear losses and parametric gain, which stabilizes their amplitude \cite{chembo4,kippenberg4}. The threshold excitation power for the onset of DKSs is proportional to the squared photon loss rate. This makes microring resonators one of the most employed platforms for comb generation, as ultra-high quality factors are easily achieved with almost no geometry optimization effort. In particular, crystalline and silicon-nitride (Si$_3$N$_4$) rings have achieved $Q$-factors in the $10^9$ and $10^7$ ranges, respectively, leading to threshold powers in the milliwatt and sub-milliwat regimes \cite{kippenberg1,lipson,herr2,lipson2}. While Si$_3$N$_4$ has become the standard platform for DKS generation in silicon photonics, the Kerr non-linearity of this material is relatively small, thus requiring such ultra-low loss resonances for low-power operation. 

Silicon is characterized by a Kerr coefficient ten times larger than that of Si$_3$N$_4$, and silicon ring resonators are compliant with CMOS technology. For these reasons, silicon has also been investigated for the generation of DKS in the mid-infrared \cite{gaeta6}. However, efficient DKS generation has not been demonstrated so far in the telecommunication band, due to the considerable non-linear losses occurring within this frequency range -- particularly two-photon absorption (TPA) and a variety of free-carrier effects at high excitation powers \cite{gaeta4}. Additionally, silicon ring resonators are also subject to large propagation losses, stemming from sidewall surface roughness, which set an obstacle to the achievement of ultra-high quality factors needed for low-power operation \cite{baets,lipson}.

\begin{figure}[t!]
\centering\includegraphics[width=0.45\textwidth]{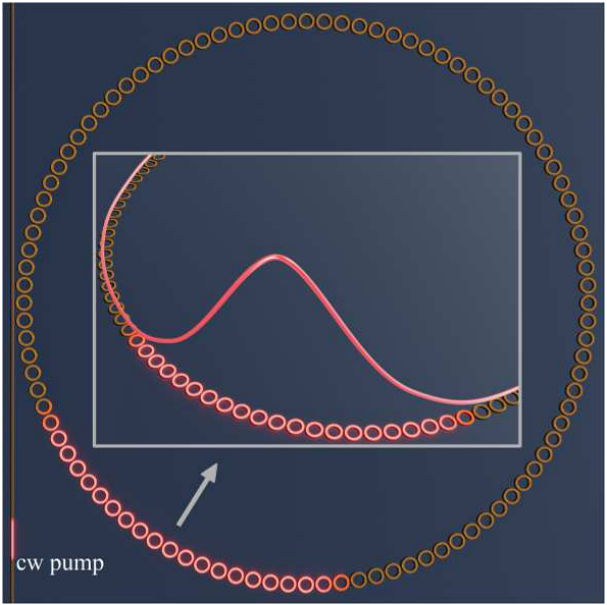}
\caption{Schematic illustration of the coupled-resonator optical waveguide formed by a closed loop of silicon microring resonators studied in this work. The system is pumped by a continuous-wave (cw) laser source in the telecom band via a bus waveguide, and the whole structure is encapsulated in silica. A dissipative Kerr soliton is depicted in red.}\label{fig_scheme}
\end{figure}

In this work, we propose a different approach to the generation of low threshold frequency combs in silicon at the telecom band. Our approach leverages structural slow light to enhance the non-linear processes, thereby requiring significantly lower values of the quality factor, which can be easily achieved with silicon microring structures. The system that we propose is illustrated in Fig.~\ref{fig_scheme}. It consists in a silica-encapsulated (SiO$_2$) coupled-resonator optical waveguide (CROW) formed by coupled single-mode silicon microrings. This configuration takes advantage of structural slow-light \cite{boyd2,momchil1,mohamed} to effectively enhance the non-linearity of the material and consequently decrease the threshold power required to trigger cascaded FWM \cite{sipe,fan,blair,derossi,notomi2,krauss2}. We study this design both in terms of first-principle FDTD simulations and using a coupled-mode effective model that has proven extremely accurate for this kind of geometries \cite{momchil1,mohamed}. The proposed Si/SiO$_2$ CROW is found to support frequency combs and DKSs with pump power in the milliwatt range at telecom frequencies, with repetition rates as low as $3.2$~GHz and a small footprint of about $0.1$~mm$^2$. We also investigate the effects of disorder, which are modeled both in terms of a reduced quality factor and by assuming randomly distributed resonant frequencies for the CROW modes. We find that DKS states are still possible in presence of disorder with standard deviation up to 1/16 of the CROW FSR, which is 20 times larger than the typical fluctuations found in standard state-of-the-art microring resonators. Our analysis opens the way to the realization of low-power DKS silicon devices operating at the telecommunication band.

The paper is organized as follows. In Sec.~\ref{formalism}, we survey the coupled-mode formalism, originally derived in Ref.~\cite{vasco2}. We study the dispersion and intrinsic losses of the device in Sec.~\ref{model}. In Sec.~\ref{FC_DKS}, we compute the non-linear dynamics of the system to find the frequency comb and DKS solutions. The effect of disorder on the DKS states are analyzed in Sec.~\ref{FC_DKS}. The conclusions of this work are drawn in Sec.~\ref{conclusions}.

\section{Formalism}\label{formalism}

The set of equations describing the non-linear dynamics of the Bloch mode slowly-varying envelopes ${\cal B}_{\alpha}(t)$ in a system of coupled Kerr resonators is given by \cite{vasco2}
\begin{align}\label{cme}
  \dot{{\cal B}}_{\alpha}(t) =& - \left[\frac{\gamma_\alpha}{2}-i\sigma_\alpha\right]{\cal B}_{\alpha}(t) \nonumber\\
  &+iG_{\alpha_0}\sum_{\mu\eta}{\cal B}_{\mu}(t){\cal B}_{\eta}^\ast(t){\cal B}_{\alpha+\eta-\mu}(t) 
  +\frac{\gamma_\alpha}{2}{\cal F}_\alpha\delta_{\alpha,\alpha_0},
\end{align}
where $\gamma_\alpha$ is the total loss rate of the Bloch mode with momentum $\alpha$, $\sigma_\alpha=\Omega_0-\omega_\alpha$ is the detuning between the laser and mode frequencies, $G_{\alpha_0}$ is the non-linear gain at the pumped mode frequency $\omega_{\alpha_0}$ and ${\cal F}_\alpha$ is the pump amplitude. The rotating frame has been introduced in Eq.~(\ref{cme}) setting ${\cal B}_\alpha(t)\rightarrow{\cal B}_\alpha(t)e^{i\sigma_\alpha t}$. In presence of TPA, which is the main source of non-linear losses in silicon at telecom frequencies, $G_{\alpha_0}$ is complex-valued and can be written as:
 \begin{equation}\label{complexG}
 G_{\alpha_0}=g_{\alpha_0}+ig_{\alpha_0}^{\rm TPA}=\left(\frac{a\omega_{\alpha_0}n_{g,\alpha_0}n_2}{\epsilon V_c}\right) + i\left(\frac{acn_{g,\alpha_0}\beta_{\rm TPA}}{2\epsilon V_c}\right),
\end{equation}
where $n_{g,\alpha_0}$ is the the group index, $V_c$ is the single resonator non-linear mode volume, $a$ is the lattice period, $n_2$ and $\beta_{\rm TPA}$ stand for the Kerr and TPA coefficients, respectively, of the material with dielectric constant $\epsilon$, and $c$ is the speed of light in vacuum. Eq.~(\ref{cme}) is normalized so that $|{\cal B}_{\alpha}(t)|^2$ is the instantaneous power of the corresponding Bloch mode propagating along the CROW direction. This choice differs from the one usually adopted in microring resonators, where the squared modulus of the envelope function represents the instantaneous energy in the resonator mode (or photon number if given in units of $\hbar \omega$). Power normalization is the natural choice when the group velocity differs considerably from the phase velocity \cite{derossi}, and it is therefore the most appropriate choice for the present work focusing on the slow-light regime.

The coupling to an external bus waveguide is included by assuming that the total loss rate $\gamma_\alpha$ is the sum of the intrinsic radiative decay rate of the mode $\gamma_{\alpha,{\rm int}}$ and the loss rate through the bus waveguide $\gamma_{\alpha,{\rm ext}}$, i.e., 
\begin{equation}\label{gammas}
    \gamma_\alpha = \gamma_{\alpha,{\rm int}} + \gamma_{\alpha,{\rm ext}}.
\end{equation}
We then use coupled mode theory~\cite{haus} to establish the relation between the pump amplitude in Eq.~(\ref{cme}) and the laser power $P$ 
\begin{equation}\label{FPrel}
\frac{\gamma_\alpha^2}{4}|{\cal F}_\alpha|^2 = \frac{c}{L\sqrt{\epsilon}}\gamma_{\alpha,{\rm ext}}P,
\end{equation}
where $L=Ma$ is the total length of the waveguide of $M$ resonators. Equation~(\ref{FPrel}) allows us to derive the external power threshold for frequency comb generation from the threshold of the internal field amplitude $|{\cal F}_{\alpha_0}|_{\rm th}^2(\sigma_{\alpha_0})$ previously derived in Ref~\cite{vasco2}
\begin{multline}\label{Pth}
P_{\rm th}(\sigma_{\alpha_0}) = \frac{L\sqrt{\epsilon}\gamma^2_{\alpha_0}}{8\eta cg_{\alpha_0}}f(\kappa) \left[ 1 + \frac{4\sigma_{\alpha_0}^2}{\gamma^2_{\alpha_0}} +\frac{4\sigma_{\alpha_0}}{\gamma_{\alpha_0}}f(\kappa) \right.\\ 
\left. + \frac{2g_{\alpha_0}^{\rm TPA}}{g_{\alpha_0}}f(\kappa) + \frac{|G_{\alpha_0}|^2}{g_{\alpha_0}^2}f^2(\kappa) \right].
\end{multline}
In Eq.~(\ref{Pth}), $\eta=\gamma_{\alpha_0,{\rm ext}}/\gamma_{\alpha_0}$ is the coupling efficiency, which is equal to $1/2$ for critical coupling, and $f(\kappa)=(\sqrt{1+\kappa^2}+2\kappa)/(1-3\kappa^2)$ is a function of the material properties only, with $\kappa=g_{{\alpha_0}}^{\rm TPA}/g_{{\alpha_0}}=c\beta_{\rm TPA}/(2n_2\omega_{{\alpha_0}})$. The minimum power threshold can be easily found by solving $\partial P_{\rm th}/\partial\sigma_{\alpha_0}=0$, leading to
\begin{equation}\label{Pthmin}
    P_{\rm th}^{\rm min} = \frac{\epsilon^{3/2} MV_c\gamma^2_{\alpha_0}}{8\eta\omega_{\alpha_0}cn_{g,\alpha_0}n_2}f(\kappa)\left[1 + \kappa f(\kappa)\right]^2,
\end{equation}
where we have used the definition of $g_{\alpha_0}$ from Eq.~(\ref{complexG}). 

Equations~(\ref{Pth}) and (\ref{Pthmin}) express the power threshold for comb generation in CROWs, in presence of TPA and slow-light. The latter is especially important due to inverse dependence of $P_{\rm th}$ on the group index $n_{g}$, which effectively enhances the Kerr non-linearity of the material and decreases the minimum power to trigger parametric FWM between the resonator modes. For typical microring resonators, the CROW system is replaced by a homogeneous waveguide, resulting in a group index which approaches the refractive index of the material, i.e., $n_{g,\alpha_0}\rightarrow\sqrt{\epsilon}$, and a total effective mode volume given by $V_{\rm eff}=MV_c$. If we further assume $f(\kappa=0)=1$ (i.e. no TPA limit), we recover from Eq.~(\ref{Pthmin}) the well known expression for the power threshold widely used for microring frequency combs \cite{kippenberg1,chembo2}
\begin{equation}\label{PminnoTPA}
    P_{\rm th}^{\rm min} (n_{g,\alpha_0}\rightarrow\sqrt{\epsilon},\beta_{\rm TPA}\rightarrow0) = \frac{\epsilon V_{\rm eff}\gamma^2_{\alpha_0}}{8\eta\omega_{\alpha_0}cn_2},
\end{equation}

\begin{figure}[t!]
\centering\includegraphics[width=0.45\textwidth]{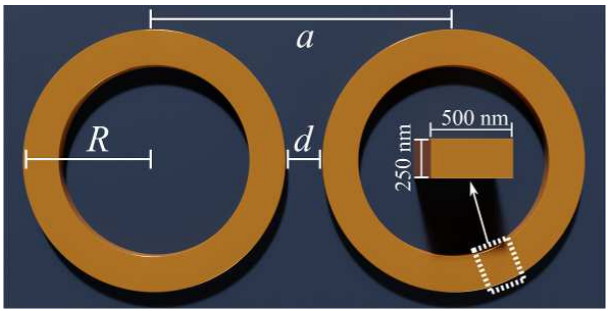}
\caption{System of two coupled silicon ring resonators with radius $R=5~\mu$m and cross section $500\times250$~nm$^2$, separated by a distance $d$. The center-to-center separation is given by $a=2R+d+500$~nm and the structure is completely encapsulated in silica.}\label{fig_crings}
\end{figure}

\begin{figure*}[t!]
\centering\includegraphics[width=0.95\textwidth]{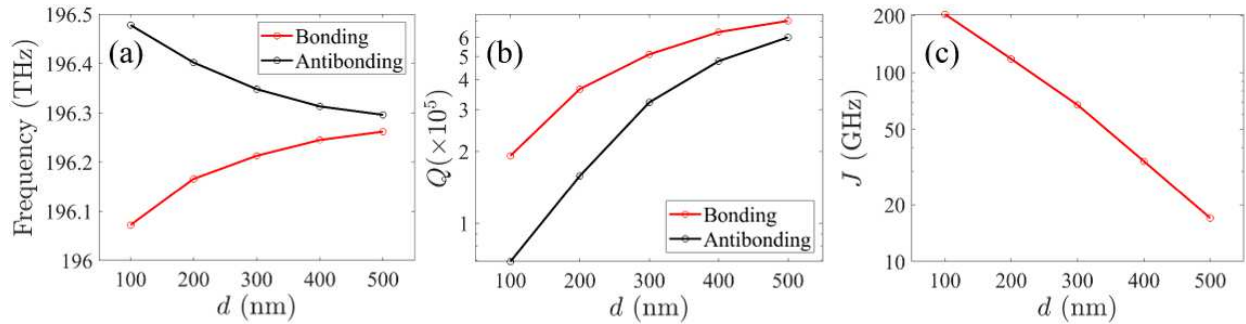}
\caption{(a) Normal mode frequencies of the bonding and antibonding states arising from the coupling of two identical single mode ring resonators. (b) Quality factors of the modes in (a). (c) Coupling strength between the microring modes, defined as half the normal-mode frequency separation from panel (a).}\label{fig_lumcrings}
\end{figure*}

\begin{figure}[t!]
\centering\includegraphics[width=0.45\textwidth]{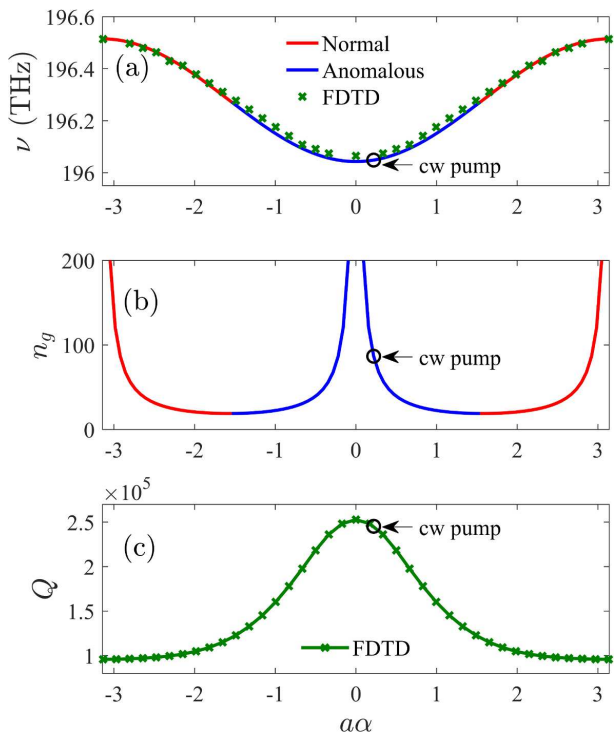}
\caption{(a) Tight-binding dispersion of the CROW. Regions where dispersion is normal and anomalous are colored in red and blue, respectively. {The corresponding FDTD simulations with Bloch boundary conditions are also shown (green crosses).} (b) Group index of the modes in (a). {(c) Simulated $Q$ factor of the band in (a)}. A cw pump is applied at $a\alpha_0=0.22$.}\label{fig_dispersion}
\end{figure}

\section{System and model} \label{model}

Whispering-gallery modes are efficiently confined within dielectric rings due to total internal reflection. The evanescent field immediately outside the ring typically decays over a short distance compared to the ring size, even when the refractive index contrast of the core-cladding is not large. This strong light confinement allows us to employ a tight-binding (TB) model in the weak and nearest-neighbor approximation to accurately describe our CROW system depicted in Fig.~\ref{fig_scheme}. The TB dispersion relation is analytical and given by \cite{sphicopoulos} 
\begin{equation}\label{TBdis}
    \omega(\alpha) = \omega_c - 2J\cos(a\alpha), 
\end{equation}
with $w_c$ denoting the mode frequency of the single rings and $J$ the coupling strength between two rings. Eq.~(\ref{TBdis}) holds under the assumptions of single mode resonators and periodic boundary conditions. The former is fulfilled as long as the FSR of the microring is much larger than the TB bandwidth $2J$, which is easily achieved for small-size rings as discussed below. The latter applies to the closed loop of coupled rings, as shown in Fig.~\ref{fig_scheme}, resulting in negligible bending losses \cite{yariv}. In order to compute the model parameters $\omega_c$ and $J$, we consider the system of two identical rings separated by a distance $d$, as illustrated in Fig.~\ref{fig_crings}. The rings have radius $R=5~\mu$m, cross section $500\times250$~nm$^2$ and refractive index $n_{\rm si}=3.47$ (silicon at telecom frequencies) \cite{minghao}. Additionally, the whole system is assumed to be encapsulated in silica (SiO$_2$). Figure~\ref{fig_lumcrings} shows the results obtained from first-principles FDTD calculations carried out with a commercial software \cite{lumerical}. The normal modes of the coupled rings, arising from the TE whispering-gallery mode at $w_c/2\pi=196.27$~THz in each ring, are shown in Fig.~\ref{fig_lumcrings}(a). The fields of the bonding and anti-bonding modes are respectively even and odd under inversion with respect to the center of the system. We report in Fig.~\ref{fig_lumcrings}(b) the associated quality factors, which for large distance $d$ approach value of the uncoupled system, $\sim 7.8\times10^5$. The coupling strength $J$, shown in Fig.~\ref{fig_lumcrings}(c), is extracted from  Fig.~\ref{fig_lumcrings}(a) and is defined as half the normal-mode frequency separation. From FDTD simulations of the single resonator, we obtain a ring FSR of $2.12$~THz at the frequency $w_c/2\pi$, i.e. roughly five times larger than the largest value of $2J$ considered, thus validating our assumption of single-mode coupling. We set the distance $d=200$~nm. This value is a good compromise between large coupling strength, low intrinsic normal mode losses and fabrication feasibility \cite{baets}. For this distance, we infer from Fig.~\ref{fig_lumcrings} $J=118$~GHz, $Q^{b}=3.64\times10^5$ (bonding) and $Q^{a}=1.57\times10^5$ (anti-bonding). We plot in Figs.~\ref{fig_dispersion}(a) and \ref{fig_dispersion}(b) the corresponding dispersion relation $\nu_\alpha = \omega_\alpha/2\pi$ [Eq.~(\ref{TBdis})] and group index $n_g$, where $a=10.7~\mu$m. As depicted in Fig.~\ref{fig_scheme}, a total number of $M=100$ microrings are considered. The resulting super-ring has radius $\sim 170~\mu$m, the band being considered contains 100 Bloch modes. {We computed the dispersion relation of the super-ring using FDTD. For this calculation we assumed an elementary computational cell containing a single ring and Bloch boundary condition along the main CROW axis. The computed dispersion is displayed in Fig.~\ref{fig_dispersion}(a). The agreement with the TB result validates our nearest-neighbor coupling model. The quality factor of the band, extracted from these FDTD simulations, is reported in Fig.~\ref{fig_dispersion}(c) and ranges from $0.95\times10^5$ at the edges of the Brillouin zone, to $2.52\times10^5$ close to $a\alpha=0$}. In order to achieve stimulated parametric FWM between the CROW modes, assisted by slow-light enhancement of the non-linear response, the system is driven at $a\alpha_0=0.22$ where $n_{g,\alpha_0}=86.7$ {and $Q_{\alpha_0}=2.45\times10^5$}. We compute the minimum threshold power from Eq.~(\ref{Pthmin}) and obtain {$P_{\rm th}^{\rm min}=9.4~$mW}. For this calculation we used $V_c=4.83~\mu$m$^3$ obtained from FDTD calculations, {$\gamma_{\alpha_0,\rm int}=\omega_{\alpha_0}/Q_{\alpha_0}$}, critical coupling $\eta=1/2$ to an external bus waveguide, and non-linear coefficients of silicon at telecom frequencies \cite{jalali}, i.e., $n2=5.52\times10^{-18}$~m$^2$/W and $\beta=1\times10^{-11}$~m/W. This threshold value is comparable to the one of rings of size of several hundred micron, operating in the ultra-low loss regimes (quality factors of the order of $10^7$) \cite{kippenberg4}.    

The present CROW does not require ultra-high quality factors to generate parametric FWM at low threshold power, thanks to the structural slow-light factor $1/n_g$ entering Eqs.~(\ref{Pth}) and (\ref{Pthmin}). Ultra-high quality factors, typically achieved in Si$_3$N$_4$ resonators, would be extremely challenging to realize in silicon rings due to propagation losses originating from surface roughness \cite{jalali,lipson}. The device that we propose thus enables silicon as a material for the realization of comb devices. In what follows, we investigate the generation of frequency combs and DKS.

\begin{figure*}[t!]
\centering\includegraphics[width=0.95\textwidth]{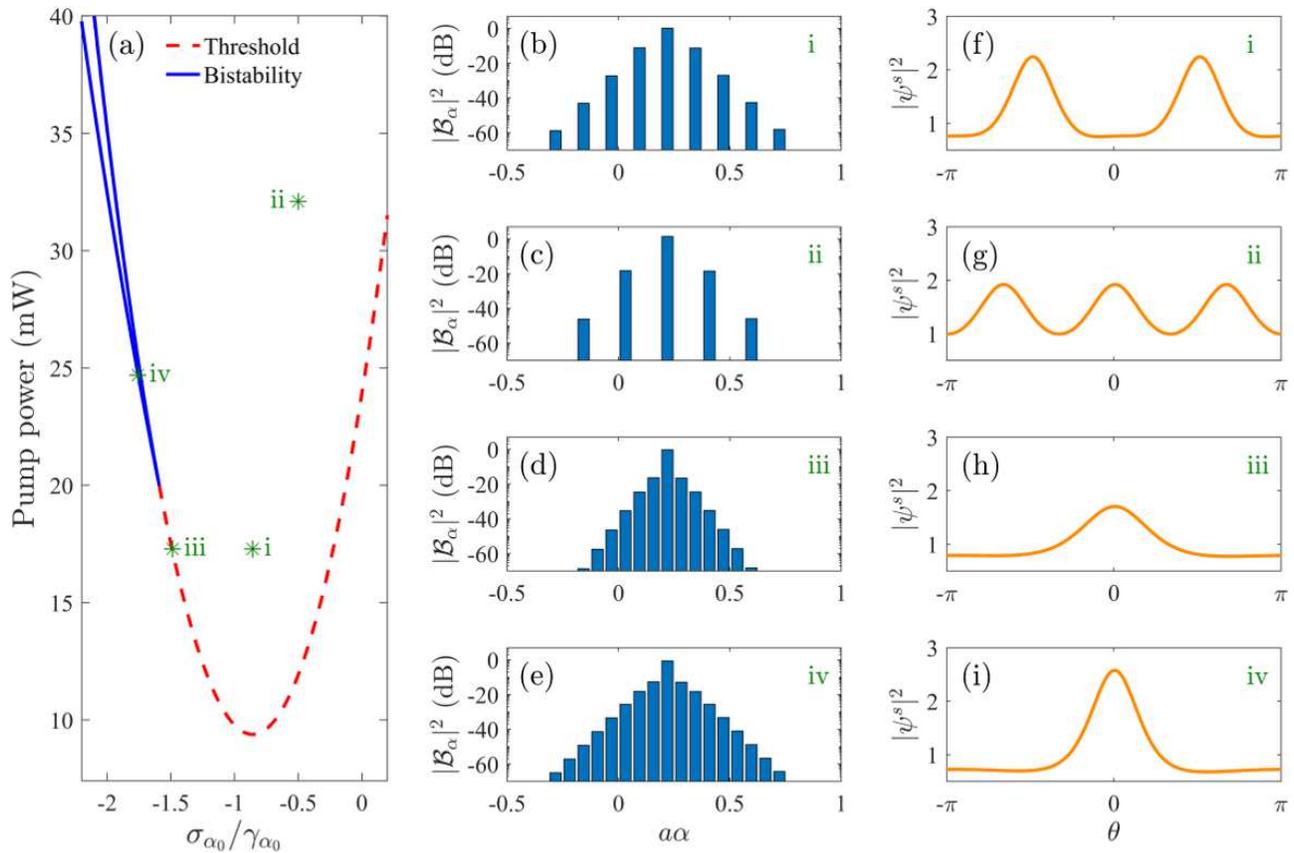}
\caption{(a) Pump power threshold (dashed-red) and bistability boundaries (continuous-blue) as a function of the laser detuning $\sigma_{\alpha_0}$ with respect to the frequency of the driven mode $\omega_{\alpha_0}$. (b) Super-critical Turing pattern of 2-FSR repetition rate ($6.4$~GHz), for a driving power {$P=17.3$~mW}. (c) Super-critical Turing pattern of {3-FSR} repetition rate ($6.4$~GHz) pumped with {$P=32.1$~mW}. (d) Soliton pulse with single FSR repetition rate ($3.2$~GHz) pumped with {$P=17.3$~mW}. (e) Soliton pulse with single FSR repetition rate ($3.2$~GHz) pumped with {$P=24.7$~mW}. (f)-(i) corresponding envelope functions of the frequency combs in (b)-(e). All $|{\cal B}_\alpha|^2$ and $|\psi^s|^2$ quantities are given in units of $\gamma_{\alpha_0}f(\kappa)/(2g_{\alpha_0})$.}\label{fig_combs}
\end{figure*}

\section{Slow-light frequency combs and DKS}\label{FC_DKS}

We plot in Fig.~\ref{fig_combs}(a) the power threshold computed from Eq.~(\ref{Pth}) as a function of $\sigma_{\alpha_0}/\gamma_{\alpha_0}$ (dashed red), and the boundaries of the optical bistability region for the driven mode (continuous blue)  \cite{chembo4,kippenberg1,vasco2}. We search for the steady state solutions of Eq.~(\ref{cme}) by scanning the value of $\sigma_{\alpha_0}$ for different values of the pump power. The coupled-mode equations are integrated using an adaptive time step Runge-Kutta method and fast Fourier transform to efficiently compute the non-linear term \cite{hansson}. {We take into account the $\alpha$-dependent quality factor $Q_\alpha$ from Fig.~\ref{fig_dispersion}(c), with corresponding intrinsic loss rates $\gamma_{\alpha,\rm int}=\omega_{\alpha}/Q_\alpha$} and total loss rates given by $\gamma_\alpha=2\gamma_{\alpha,\rm int}$ at critical coupling. The pump field at time $t=0$ is assumed to have Gaussian shape $\psi(\theta,0)=\exp[-0.5( M\theta/2\pi)^2]$ along the CROW, where $\psi(\theta,t)=\sum_\alpha{\cal B}_{\alpha}(t)e^{-i(\alpha-\alpha_0)\theta L/2\pi}$ is the envelope function of the solution and $\theta$ is the polar angle denoting the position along the CROW. We show in Figs.~\ref{fig_combs}(b)-\ref{fig_combs}(e) four representative frequency combs found for the four values of pump power and detuning highlighted in Fig.~\ref{fig_combs}(a). Figures~\ref{fig_combs}(b)-\ref{fig_combs}(c) correspond to super-critical Turing patters (they are excited above threshold) with 2-FSR and 3-FSR spacing, respectively. Fig.~\ref{fig_combs}(d)-\ref{fig_combs}(e) show sub-critical combs (excited below threshold) with single FSR spacing, which are the signature of soliton structures. The {steady state} envelope functions of these combs, {denoted as $\psi^s(\theta)$}, are shown in Figs.~\ref{fig_combs}(f)-\ref{fig_combs}(i). Two and three Turing rolls emerge  respectively for the spectra with 2-FSR and 3-FSR spacing, in agreement with the Lugiato-Lefever model predictions \cite{chembo4}, while single solitons arise in the cases of 1-FSR spacing. 

Even in presence of TPA, the present silicon CROW supports DKS structures at telecom frequencies and low power. This is achieved thanks to the non-linear enhancement provided by slow-light, which enable operating at significantly lower $Q$-values than those required in current microring resonators. The advantage brought by slow-light comes at the expense of the comb band-width, which is of the order of $45$~GHz with a repetition rate of $3.2$~GHz. Nevertheless, few-GHz repetition rates are desirable for applications in spectroscopy and signal processing in electronics \cite{diddams}, and they are usually obtained in centimeter-size resonators \cite{vahala} -- much larger than the $340~\mu$m diameter of the present device.

\begin{figure*}[t!]
\centering\includegraphics[width=0.95\textwidth]{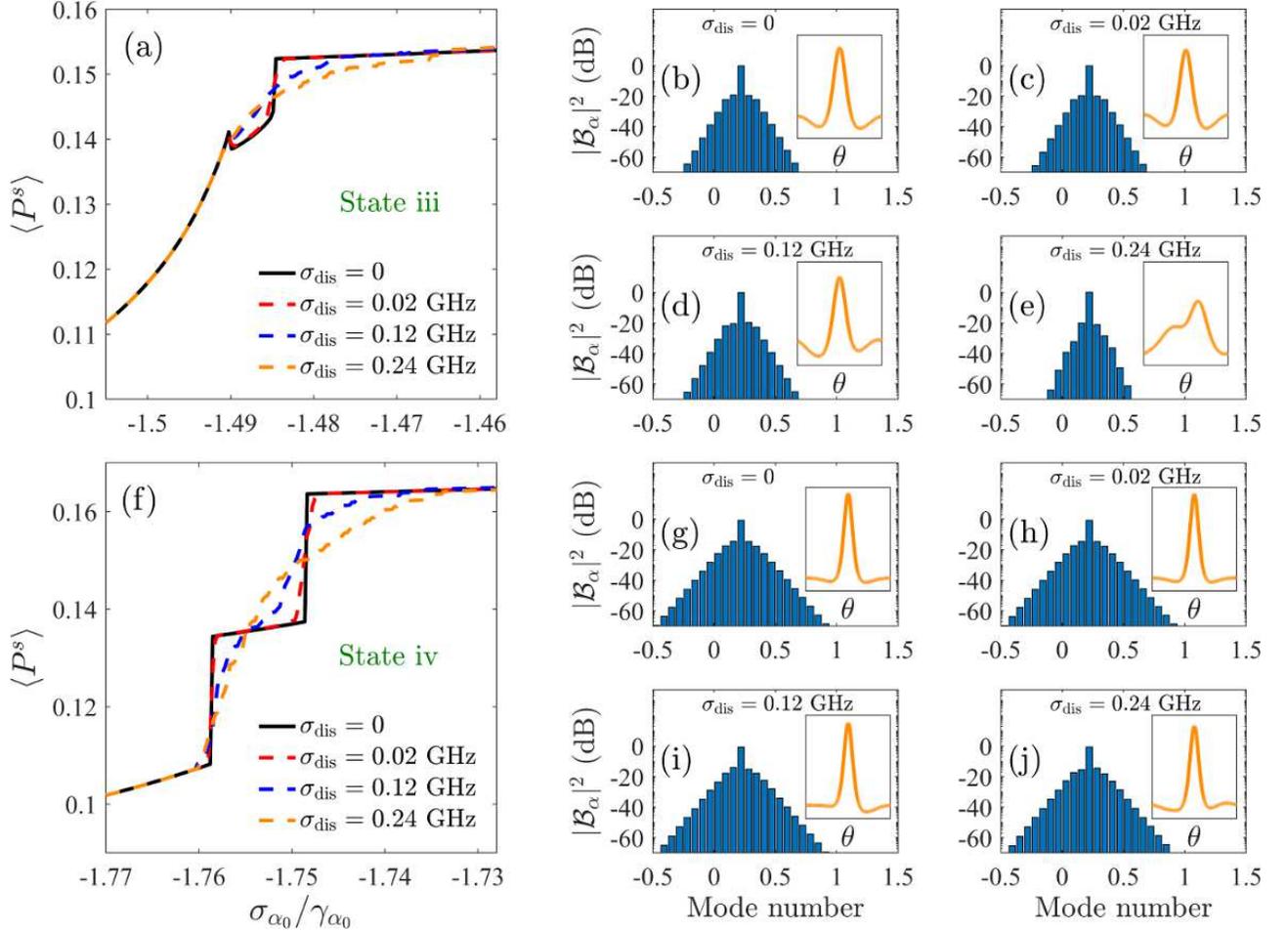}
\caption{(a) {Averaged $P^s$} over 50 disorder realizations for different disorder magnitudes $\sigma_{\rm dis}$, as a function of the laser detuning $\sigma_{\alpha_0}$. The pump power is set to {$P=54$~mW}. (b) Frequency comb of the soliton state computed at {$\sigma_{\alpha_0}/\gamma_{\alpha_0}=-1.4874$} for $\sigma_{\rm dis}=0$ (state iii). The envelope function $\psi(\theta)$ is displayed in the inset. (c)-(e) Selected disorder realizations of the soliton combs computed at {$\sigma_{\alpha_0}/\gamma_{\alpha_0}=-1.4874$} for the different disorder magnitudes considered in (a). (f) Same as (a) with a pump power of {$P=77.1$~mW}. (g)-(j) Same as (b)-(e), computed at {$\sigma_{\alpha_0}/\gamma_{\alpha_0}=-1.75355$} (state iv). All $|{\cal B}_\alpha|^2$ and $|\psi^s|^2$ quantities are given in units of $\gamma_{\alpha_0}f(\kappa)/(2g_{\alpha_0})$.}\label{fig_dis}
\end{figure*}

\section{Effects of disorder}\label{disorder}

In order to study the robustness of our results against structural disorder arising at the fabrication stage, we model the effects of surface roughness by assuming a random deviation of the Bloch mode frequencies from the ideal values, and an additional loss channel. The quality factor associated to this additional loss channel is denoted by $Q_r$. The total quality factor of the ring mode, denoted as $Q_{c,\rm tot}$, is then computed as 
\begin{equation}\label{Qctot}
    \frac{1}{Q_{c,\rm tot}} = \frac{1}{Q_c} + \frac{1}{Q_r},
\end{equation}
with $Q_c$ representing the FDTD-computed quality factor of the ideal system with perfect walls. As silica encapsulation smooths out structural imperfections at the ring surfaces \cite{painter,vasco3}, a lower bound to the $Q_r$ can be estimated from a silicon-on-insulator (SOI) configuration (where sidewalls are exposed directly to air). The total quality factor of the same ring geometry but in a SOI setup is reported in Ref.~\cite{minghao} with a value of $Q^{\rm SOI}_{c,\rm tot}=2.2\times10^5$. We then carry out the corresponding SOI simulation from which we obtain $Q^{\rm SOI}_{c}=7\times10^5$. Using $Q^{\rm SOI}_{c,\rm tot}$ and $Q^{\rm SOI}_{c}$ in Eq.~(\ref{Qctot}) we get $Q_r^{\rm min}=3.2\times10^5$. Thereby, the normal mode quality factors of the CROW in presence of surface roughness, {$Q_{\alpha,r}$}, can be approximated by assuming the lower bound of $Q_r$, i.e.
\begin{equation}\label{Qbtot}
    \frac{1}{Q_{\alpha,r}} = \frac{1}{Q_\alpha} + \frac{1}{Q_r^{\rm min}}.
\end{equation}
This increases the minimum power threshold of Eq.~(\ref{Pthmin}) from {$P_{\rm th}^{\rm min}=9.4$~mW to $P_{\rm th}^{\rm min}=29.3$~mW}, under critical coupling conditions ($\eta=1/2$). 

The effect of disorder on the frequencies of the Bloch modes is modeled as
\begin{equation}\label{wdis}
    \omega^{\rm dis}_\alpha = \omega_\alpha+\delta\omega,
\end{equation}
where $\delta\omega$ are random fluctuations following a Gaussian probability distribution with standard deviation $\sigma_{\rm dis}$. This random term models random variations of both $\omega_c$ (due to sidewall roughness) and $J$ (due to imperfect positioning of the microring resonators). The new parameters {$\gamma^r_{\alpha,\rm int}=\omega_{\alpha_0}/Q_{\alpha,r}$} and $\omega^{\rm dis}_\alpha$ are introduced in the coupled-mode equations Eq.~(\ref{cme}) which is solved under the same conditions as in Sec.~\ref{FC_DKS}. We first focus on the soliton state of Fig.~\ref{fig_combs}(d) and plot in Fig.~\ref{fig_dis}(a) {the average of $P^s=\int_{-\pi}^\pi|\psi^s(\theta)|^2d\theta/2\pi$} over an ensemble of 50 disorder realizations as a function of the laser detuning $\sigma_{\alpha_0}$, for different values of $\sigma_{\rm dis}$. Notice that, for perfectly ordered frequencies, i.e., $\sigma_{\rm dis}=0$, where the effect of surface roughness is only considered on the quality factor of the CROW modes {$Q_{\alpha,r}$}, the characteristic step in {$P^s$} (signature of the soliton appearance) is still present but with a pump power increased from {$P=17.3$~mW to $P=54$~mW}. We show in Fig.~\ref{fig_dis}(b) the frequency comb and corresponding envelope function in the inset along the super-ring polar angle $\theta$ at {$\sigma_{\alpha_0}/\gamma_{\alpha_0}=-1.4874$} [same of Fig.~\ref{fig_combs}(d)]. As disorder in the Bloch mode frequencies increases, the step in the averaged waveguide power becomes less evident thus making soliton states more unlikely to appear. Selected realizations for the different disorder magnitudes employed in this analysis are shown in Figs~\ref{fig_dis}(c)-\ref{fig_dis}(e). Here, we clearly see the effects of disorder on the soliton envelope, which starts to loss its spatial localization at $\sigma_{\rm dis}=0.24$~GHz. We now focus on the state shown in Fig.~\ref{fig_combs}(e), whose disorder analysis is presented in Fig.~\ref{fig_dis}(f). In order to recover the same step features in presence of sidewall roughness, the pump power is increased from {$P=24.7$~mW to $P=77.1$~mW}. Selected realizations are also shown for this case in Figs.~\ref{fig_dis}(g)-\ref{fig_dis}(j) at {$\sigma_{\alpha_0}/\gamma_{\alpha_0}=-1.75355$} [same of Fig.~\ref{fig_combs}(e)]. Interestingly, for the largest disorder magnitude considered, the step in {$\langle P^s\rangle$} is still present and a spatially localized soliton envelope is more likely to appear than in Fig.~\ref{fig_dis}(e), thus making this higher-power state more robust against random fluctuations on $\omega_\alpha$.

While disorder magnitudes of the order of $0.2$~GHz might be challenging to achieve in modern fabrication techniques, they are $\sim 1/16$ of the CROW FSR ($3.2$~GHz), which is around 23 times the typical relative frequency fluctuation (with respect to the FSR) in modern microring resonators \cite{kippenberg2}. Therefore, our results clearly evidence the extremely robust nature of DKS against random fluctuations on the resonator frequencies.  

\section{Conclusions}\label{conclusions}

We have studied the formation of frequency combs and DKS in a CROW made of silica-encapsulated ring resonators, operating at telecommunication frequencies. Thanks to slow-light enhancement of the non-linear response of the CROW, combs and DKSs can be generated for significantly lower values of the quality factor, than those typically required by a microring resonator. These low values of the quality factor can be achieved in silicon structures even in presence of losses induced by surface roughness. Our study shows that combs and DKSs can arise already for pump power in the 10~mW range. 
% THIS SENTENCE IS WRITTEN UPSIDE-DOWN!!!
%The ultra-low losses condition generally required by modern microring resonators platforms, and challenging to achieve in silicon because of the relatively high propagation losses, is relaxed in our system due to slow-light non-linear enhancement, which leads to milliwatt threshold power for comb generation in presence of TPA.
{This range of pump values also allows to rule out additional non-linear loss mechanisms due to free-carrier absorption at the 1.55~$\mu$m band, which become relevant for input intensities larger than 10~GW/m$^2$ \cite{davies}.} 

We have also addressed the effects of surface roughness and random frequency fluctuations arising from surface roughness introduced at the fabrication stage. Our results show that DKSs are robust against fabrication imperfections for values of the disorder amplitude well above those routinely achieved in microring resonators. The only effect of disoreder is a moderate increase of the power required for comb generation, due to the reduced quality factor. 

The CROW structure that we propose opens the way to efficient, compact and low power comb and DKS generation in silicon devices at telecom frequencies.

%\bibliographystyle{apsrev4-1}
%\bibliography{bibliography}

%merlin.mbs apsrev4-1.bst 2010-07-25 4.21a (PWD, AO, DPC) hacked
%Control: key (0)
%Control: author (0) dotless jnrlst
%Control: editor formatted (1) identically to author
%Control: production of article title (0) allowed
%Control: page (1) range
%Control: year (0) verbatim
%Control: production of eprint (0) enabled
%

\end{document}